\documentclass[9pt]{article}
\usepackage[utf8]{inputenc}
\usepackage{multicol}
\usepackage{geometry}
\usepackage{graphicx}
\usepackage{wrapfig}

\usepackage{upgreek}
\usepackage{color}
\usepackage{amsmath}
\usepackage{amssymb}

\usepackage{natbib}
\setcitestyle{numbers,open={[},close={]}}

\newcommand{\rmd}{\mathrm{d}}

\title{\textsc{Stability of respiratory-like droplets under evaporation}}
\author{Carola Seyfert\textsuperscript{a}, 
        Javier Rodr\'{i}guez-Rodr\'{i}guez\textsuperscript{\dag b},
        Detlef Lohse\textsuperscript{a,c},
        Alvaro Marin\textsuperscript{a},
        }

\date{\today}

\begin{document}

\maketitle

\vspace*{-0.8cm}

\begin{center}
\textsuperscript{a} Physics of Fluids Group, Department of Science and Technology, Mesa+ Institute, Max Planck Center for Complex Fluid Dynamics and J. M. Burgers Centre for Fluid Dynamics, University of Twente, 7500 AE Enschede, The Netherlands

\textsuperscript{b} Departamento de Ingenier\'{i}a T\'{e}rmica y de Fluidos, Universidad Carlos III de Madrid, 28911 Leganes, Spain

\textsuperscript{c} Max Planck Institute for Dynamics and Self-Organization, 37077 G\"{o}ttingen, Germany
\end{center}

\vspace*{-0.1cm}

\begin{multicols}{2}
[
\begin{center}
\section*{Abstract}
\end{center}
Pathogens contained in airborne respiratory droplets have been seen to remain infectious for periods of time that depend on the ambient temperature and humidity. In particular, regarding the humidity, the empirically least favorable conditions for the survival of viral pathogens are found at intermediate humidities. However, the precise physico-chemical mechanisms that generate such least-favorable conditions are not understood yet.
In this work, we analyze the evaporation dynamics of respiratory-like droplets in air, semi-levitating them on superhydrophobic substrates with minimal solid-liquid contact area.
Our results reveal that, compared to pure water droplets, the salt dissolved in the droplets can significantly change the evaporation behaviour, especially for high humidities close to and above the deliquesence limit. Due to the hygroscopic properties of salt, water evaporation is inhibited once the salt concentration reaches a critical value that depends on the relative humidity. The salt concentration in a stable droplet reaches its maximum at around 75\% relative humidity, generating conditions that might shorten the time in which pathogens remain infectious.
]

\vspace*{-0.2cm}

\section*{Introduction}
The recent COVID19 pandemic has revealed how little we know about the physical mechanisms that allow infectious agents to be transmitted through air. Indeed, airborne pathogens like influenza viruses and coronaviruses, but also bacterial agents like tuberculosis or legionellosis, can survive for hours or even days in exhaled droplets \cite{Geller2012, Tang2009reviewAirbornePathogens}.

For instance, Influenza, an enveloped virus like the SARS-CoV-2, is typically transmitted more efficiently during the winter season (typically cold and dry), but also in rainy seasons in warmer climates (warm and wet) \cite{tamerius2013influenza,Bov2021Humidity}. However, transmission efficiency is a very complex issue since it involves both physical aspects like the environmental conditions on the transmission route (air flows, humidities and temperatures), as well as biological ones, such as for example the host's or recipient's immune system reaction to these environmental conditions.
During airborne transmission, the virus is usually expelled from a host in aerosolized droplets, through coughing, speaking, laughing, or simply breathing \cite{Morawska2009,bourouiba2014violent,Bourouiba2021,morawska2020time,asadi2020coronavirus}. These water-based respiratory droplets contain, besides the virus, a wide variety of salts, proteins and surfactants, the concentrations of which determine the salinity and pH of the solution.
% Biological components in these droplets besides the virus, such as salts, proteins and water, determine the salinity and pH of the solution, and can of course present with varying concentrations.
Previous empirical studies have shown that the relative humidity plays a significant role in the survival of {airborne} influenza viruses \cite{kormuth2018influenza}. Assuming that a pathogen can survive longer in a water-based suspension at the right pH and salinity, one would expect that very dry conditions would certainly be very adverse for a pathogen. This seems to be the case for most bacteria, with only few exceptions \cite{Survival2020Marr}.

However, recent data have shown that viruses do not only survive at high relative humidities but, contrary to intuition, also at extremely low ones, with a substantial decay in activity for intermediate humidities \cite{morris2021elife}. The precise values of such humidities vary for different studies \cite{schaffer1976survival,benbough1971survival,vejerano2018physico}, as do the proposed mechanisms for such a non-linear dependence of the viral activity with the humidity. 
History provides cases of viruses that were preserved on purpose in dried form, for example as powders. This is the case for the Vaccinia virus, the active constituent of the vaccine that eradicated smallpox, making it the first human disease to be eradicated. The virus/vaccine started being transported from ``arm to arm'', but in the 19th century there are records of numerous doctors reporting that the virus was better preserved active when ``...wrapt in lint and secured from air, heat, and moisture, they have sometimes continued efficacious for several months'' \cite{collier1954preservation}.
Early studies in the first half of the 20th century \cite{Edward1941Survival}, recreated in different experimental works more recently \cite{thomas2008survival,asadi2020influenza}, have confirmed that also Influenza virus can be transmitted through dried nuclei. 

These facts indicate that, in order to have better control of airborne infections via viral vectors in the future, we must gain an understanding of the physical processes occurring within complex respiratory evaporating droplets. {This will allow us to propose mechanisms for the survival of enveloped viral pathogens, based on the conditions that the virions must withstand during drop evaporation under controlled environmental humidity and temperature}.

In this work we analyze the evaporation process of spherically-shaped droplets on top of a superhydrophobic substrate with fractal-like microstructures \cite{Berenschot2013fractalfabrication}. The droplets are water-based solutions containing salt, mucin and surfactants, mimicking the content of respiratory droplets. 
By minimizing the contact of the liquid phase with the substrate, we are able to emulate the drying of exhaled aerosol droplets in a much more controlled way than in aerosol chambers. At the same time it allows us to retrieve the intact dry remains for further analysis using scanning electron microscopy.

We start by analyzing the evaporation of droplets with different components and at different humidities. Using diffusion-limited evaporation models, we can precisely determine the far-field vapor concentration felt by the droplets and accurately obtain the time evolution of the salt concentration within each droplet, depending on the environmental conditions. Evaporating the droplets on such substrates also allows us to study the role of mucin and surfactants in the final structure of those droplets that dry out completely on top of the substrate. We finalize the paper with the discussion of the potential implication of our results on the survival of viral pathogens in aerosolized respiratory droplets, in light of recent results from the literature. 

\begin{figure*}%[tbhp]
\centering
\includegraphics[width=2\columnwidth]{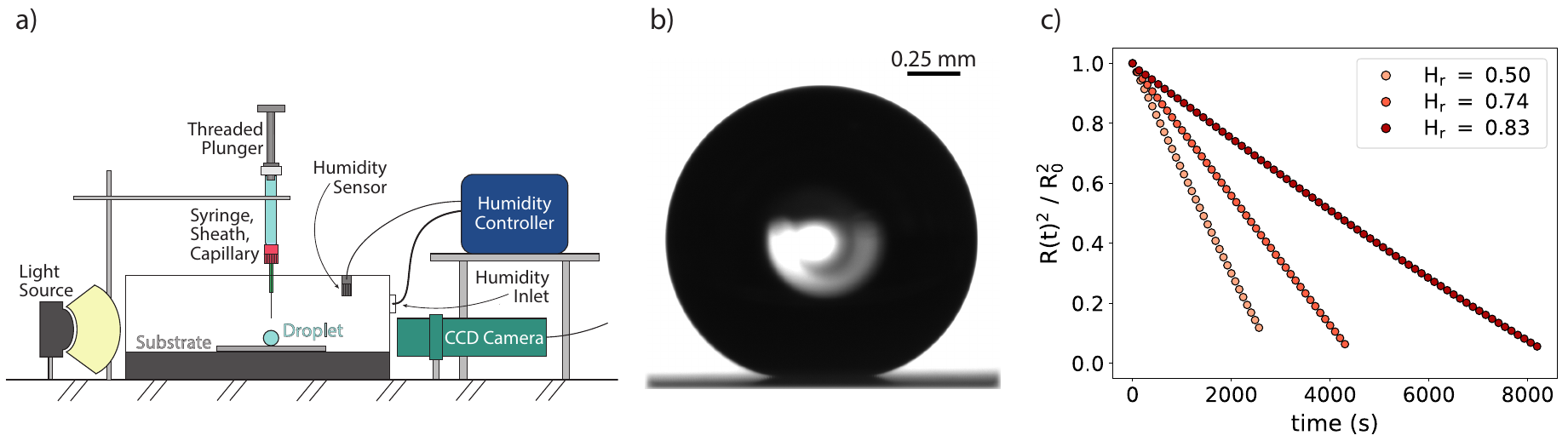}
\caption{(a) Sketch of the experimental setup employed in the study. (b) Typical side view image of a water droplet sitting on the superhydrophobic substrate. We extract contact angle and droplet volume during the evaporation from these images. (c) Example measurements for 100\% water droplets at different humidities and room temperature of 20$^\circ$C. The droplets evaporate following the classic so-called $D^2$-law. Here $R(t)$ is the radius of a volume-equivalent sphere. The droplet volume decreases, until the whole droplet has vanished. The data acquisition ends when we reach the limit of the spatial resolution of the camera. 
}
\label{fig:1}
\end{figure*}

\section*{Results}

\subsection*{Experimental results with respiratory-like droplets}

The experiments aim to mimic the evaporation process of aerosolized respiratory droplets and to analyse it in detail. This will allow us to monitor the time evolution of the salt concentration that virions might be exposed to. We evaporate droplets containing salt, protein and surfactant at determined concentrations to mimic those found in human saliva (more details in subsection \emph{Methods}\ref{subsec:materials}). The droplets evaporate on a superhydrophobic substrate which minimizes the liquid-solid contact area (more details in subsection \emph{Methods}\ref{subsec:substrate}).

\begin{figure*}
 	\centering
 	\includegraphics[width=2\columnwidth]{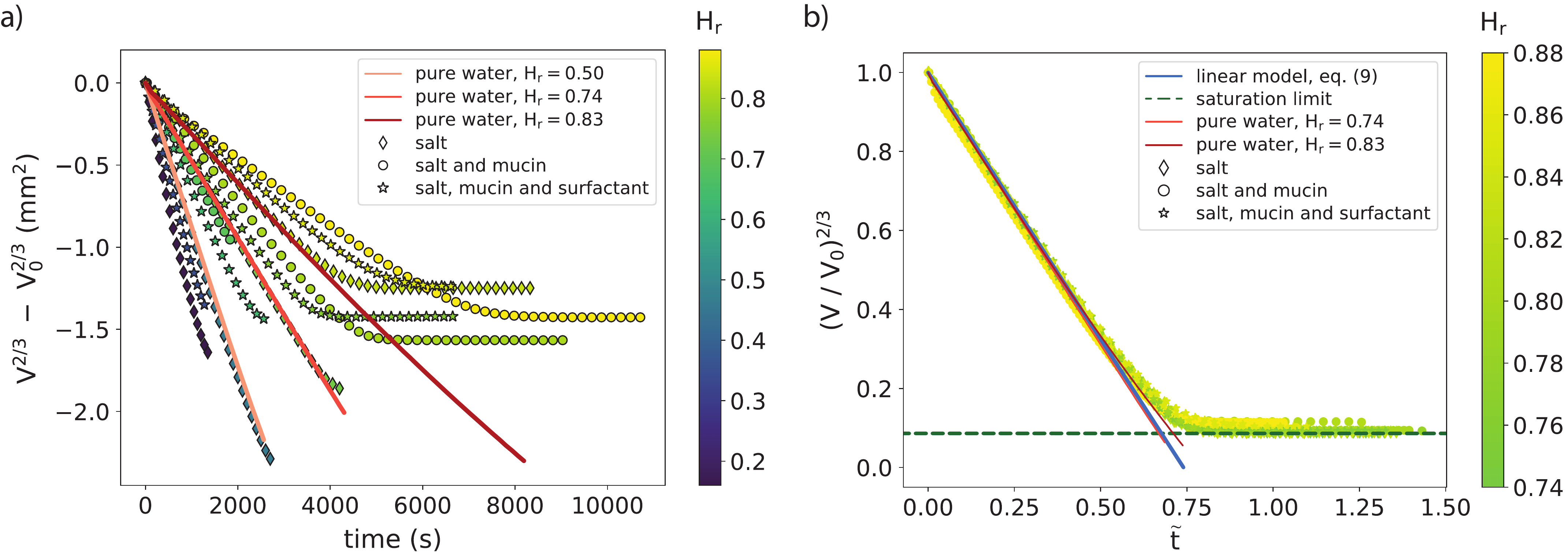}
 	\caption{{
 	(a) Droplet volume, as $V(t)^{2/3}-V_0^{2/3}$ (in mm$^2$) with ${V_0=V(t=0)}$, against time (in seconds) for different humidities (shown in a colormap, see the colorbar) and compositions (denoted in the symbols, see the legend). The wide, uniform spread of the data suggests that the evaporation process only depends on humidity and temperature, and is independent of the droplet composition at early times. In overlay, we plot the experiments of pure water droplets, also shown in Fig. \ref{fig:1}(c). In comparison, we see that the droplets containing organic components start to stabilize in volume at late times and high humidities. Due to the presence of sodium chloride, evaporation stops and the droplets keep a stable liquid volume. % and remain liquid. 
 	 This evaporation halt happens for $\rm{H_{r}~\gtrsim~0.75}$, which is the deliquesence limit for sodium chloride.
 	(b) Normalized droplet volume, as $\left(V(t)/V_0\right)^{2/3}$, against a normalized time $\rm{\widetilde{t}}$ (see eq.\ref{eq:timeScale}), for all experiments with droplets for which $\rm{H_{r}~\geqslant~0.75}$. Note the changed colorbar compared to panel (a), to denote the relative humidity of each experiment. Composition of droplets is again denoted in the legend. Using this normalization, which is independent of the droplet content, all experiments overlap. We furthermore plot two asymptotic limits of the evaporation process: the blue line denotes the linear model of volume reduction for pure water droplets (see eq.\ref{eq:water_mass_conservation_short_times} and eq.\ref{eq:timeScale}). The dashed green line indicates the prediction for the volume stability,
%  	$\left(V/V_0\right)^{2/3}$
 	based on the initial sodium chloride concentration in the droplets and the saturation concentration.}
 	}
 	\label{fig:allRH}
 \end{figure*}

Fig. \ref{fig:1}(c) shows example results of evaporation experiments at different humidities for \emph{pure} water droplets. As can be seen, the evaporation of such droplets follows a well-known so-called $D^2$-law, saying that the square of the diameter linearly decreases in time, from which we deduce that the process is diffusion-limited \cite{Popov2005}. The first question is whether respiratory-like droplets will also evaporate following such a $D^2$-law or if any of their components (sodium chloride, mucin or surfactant) would alter their behavior, one way or another. To answer this question, we perform experiments at different humidities for 3 types of solutions: (a) sodium chloride (9 g/L), (b) sodium chloride and mucin (3 g/L) and (c) sodium chloride, mucin and the surfactant DPPC (0.5 g/L). The results are shown in Fig. \ref{fig:allRH}(a), where we plot $V^{2/3}(t)-V_0^{2/3}$ against time. Here $V(t)$ is the droplet volume at a given time $t$ and $V_0 = V(0)$.

The humidity shown in this and subsequent plots has been computed from the diffusion-limited evaporation rate model (described below) and was not taken from the humidity sensor in the set-up, which is used to confirm the stability of the ambient conditions in the chamber. This way, we can be certain of the far field humidity felt by the droplet. We have observed a consistent bias between the sensor's humidity and the one computed from the evaporation-rate model, which will become relevant when we discuss our own results and compare them with other results from the literature.

In Fig. \ref{fig:allRH} we can see that, during a first stage of the evaporation process, the droplet volume evolves in time just like pure water droplets (solid lines), regardless of their composition (symbols). This is fairly well described by a simple model valid for pure water solutions, and has recently been shown for colloidal solutions \cite{seyfert2021evaporation}. Thus we conclude that the early evaporation process is not affected by the composition of the droplet, no matter the relative humidity, just as suggested by our analytical model described below.

However, remarkably, the composition of the droplet, in combination with high relative humidities, starts to play a crucial role for the evaporation dynamics at late times. Above a certain critical relative humidity, all droplets containing salt reach a stable volume. This critical relative humidity is around 75\%, and thus much smaller than 100\%.
The final liquid volume depends on the composition, most strongly on the amount of sodium chloride initially introduced in the solution. The inhibition of evaporation (and hence, volume reduction and water loss) is a result of the vapor pressure dependence on the salt concentration of the solution. As the droplet loses water, the salt concentration increases and the vapor pressure decreases.
The driving force of any evaporative process can be expressed as a difference in vapor pressures between the ambient and the liquid surface. Thus this driving force becomes smaller as the salty droplet evaporates.
As it turns out here, at 20$^{\circ}$C, the minimum equilibrium vapor pressure of a stable sodium chloride solution, attained at saturation, corresponds to the vapor pressure of pure water at a relative humidity of approximately 75\%. Consequently, all droplets containing sodium chloride at a relative humidity of 75\% or above will evaporate only until they reach a salt concentration such that its equilibrium vapor pressure matches that of pure water at that relative humidity.

The consequences for respiratory aerosols are crucial: if this droplet were a respiratory aerosolized droplet in an environment at 20 $^{\circ}$C with high humidity (above its equilibrium relative humidity of 75\%), this would result in a stable droplet aerosol, that would remain liquid indefinitely without drying. In order to understand and discuss the consequences of such a behavior, we proceed to model the system and compare it with our experimental results.

\subsection*{Analytical model}

We consider here the evaporation of a droplet containing a non-volatile solute that modifies the vapor pressure of the solution (in our case, sodium chloride dissolved in water). If the evaporation is limited by diffusion, then the mass of solvent, $m_w$, in an isolated spherical drop of radius $R$ changes with time as \cite{Xie_etalIndoorAir2007}:
\begin{equation}
    \frac{\rmd m_w}{\rmd t} = 4 \pi R D \frac{p}{{\cal{R}} T} \ln\left(\frac{p-p_{v,a}}{p-p_{v,\infty}}\right),
    \label{eq:water_mass_conservation_isolated}
\end{equation}
where $D$ the diffusivity of the vapor in air, $p$ and $T$ the air pressure and temperature respectively, and $\cal{R}$ the ideal gas constant. Moreover, $p_{v,a}$ and $p_{v,\infty}$ are the vapor pressure at the droplet's surface and far away from it, respectively.

In the case of a sessile droplet with contact radius $R_c$ and angle $\theta$, this equation must be modified as pointed out in \cite{Popov2005} to yield:
\begin{equation}
    \frac{\rmd m_w}{\rmd t} = \pi R_c f(\theta) D \frac{p}{{\cal{R}} T} \ln\left(\frac{p-p_{v,a}}{p-p_{v,\infty}}\right),
    \label{eq:water_mass_conservation_hydrophobic}
\end{equation}
where $f(\theta)$ is a function of the contact angle,
%given by \cite{Popov2005},
\begin{equation}
    f(\theta) = \frac{\sin \theta}{1 + \cos \theta} + 4\int_0^\infty \frac{1 + \cosh(2\theta\tau)}{\sinh(2\pi\tau)} \tanh[(\pi - \theta)\tau] \, \rmd\tau.
\end{equation}
Besides the contact angle which we assume roughly constant during the evaporation process  ($\theta \approx 150^{\rm o}$ in our experiments), an expression for the contact radius $R_c$ is required \cite{seyfert2021evaporation}, 
% We obtain that from our previous work \cite{seyfert2021evaporation},
\begin{equation}
    R_c = \left(\frac{3 V g(\theta)}{\pi}\right)^{1/3}.
\end{equation}
Here, $g(\theta) = \sin^3\theta/[(2+\cos\theta)(1-\cos\theta)^2]$ and $V = (m_w + m_s)/\rho(\mu)$ is the drop volume. The density of the water-salt solution $\rho(\mu)$ is taken from ref. \cite{lide2005crc}. It is a function of the \textcolor{black}{salt weight fraction of the solution},
\begin{equation}
    \mu = \frac{m_s}{m_w + m_s},
\end{equation}
with $m_s$ the mass of salt present in the drop (constant in time).

Since, even at 100\% relative humidity, the water vapor pressure is much smaller than the atmospheric one, we can simplify equation (\ref{eq:water_mass_conservation_hydrophobic}) to
\begin{equation}
    \frac{\rmd m_w}{\rmd t} = \pi R_c f(\theta) D \left(\frac{p_{v,\infty}}{{\cal{R}} T}-\frac{p_{v,a}}{{\cal{R}} T}\right).
    \label{eq:water_mass_conservation_isolated_simple}
\end{equation}
We can further simplify this equation by defining $p_{v,\infty} = H_r p_v(T)$, where $H_r$ is the relative humidity and $p_v(T)$ is the vapor pressure at a given temperature, which can be computed using Antoine's equation \cite{chong2021extended}. The vapor pressure of the water at the surface of the drop is given by $p_{v, a} = \chi_w  p_v(T)$, where $\chi_w$ is the so-called water activity. Although it may be estimated as the molar fraction of water in the dissolution, in this work we take its value from an experimental investigation reported elsewhere \cite{PapelaDunlopJCT1972}, where this activity is provided as a function of the salt weight fraction of the solution, $\chi_w(\mu)$. Denoting by $c_s = p_v(T)/{\cal{R}} T$ the water vapor concentration in air at saturation conditions, we can finally write:
\begin{equation}
    \frac{\rmd m_w}{\rmd t} = \pi R_c f(\theta) D c_s \left(H_r - \chi_w\right).
    \label{eq:water_mass_conservation_linear}
\end{equation}

An important consequence of this equation is that the drop stops dissolving, ${\rm d}m_w/{\rm d}t = 0$, when the vapor pressure at its surface becomes equal to that far away from the droplet, thus $\chi_w = H_r$. This means that respiratory droplets, which start from a water activity close to one, will not evaporate completely if the relative humidity is larger than the minimum water activity it can reach, $\chi_{w, {\rm min}} \approx 0.76$, this value corresponding to saturation conditions \cite{ApelblatKorinJCT1998}. This conclusion is valid also for isolated spherical drops. Notice that for relative humidities larger than the critical one, $H_r \gtrsim 0.75$\footnote{The precise value provided in the literature depends on the reference (see \cite{PapelaDunlopJCT1972} for a compilation). We report here a value 0.75, which is the most consistent with our own measurements and in good agreement with the ones reported.}, the droplet stops evaporating with a salt concentration smaller than the saturation one, as the condition $\chi_w = H_r$ is attained earlier in the evaporation process.

Finally, to compare the results of experiments with different relative humidity, it is convenient to define a time scale for the problem. To do so, we apply equation [\ref{eq:water_mass_conservation_linear}] to the first stage of the dissolution, where the salt concentration is small, $\mu \ll 1$. Consequently, $\chi_w \simeq 1$ and $\rho \simeq \rho_0$, with $\rho_0$ the density of pure water. In these conditions $m_w \approx \rho_0 V$ and we can write
\begin{equation}
    \frac{\rmd V}{\rmd t} = \pi \left(\frac{3 V g(\theta)}{\pi}\right)^{1/3} f(\theta) \frac{D c_s}{\rho_0} \left(H_r - \chi_w\right).
    \label{eq:water_mass_conservation_short_times}
\end{equation}
This equation can be integrated to yield
\begin{equation}
    \left(\frac{V}{V_0}\right)^{2/3} - 1 = - f(\theta) \left(\frac{g(\theta)}{2}\right)^{1/3} \frac{t}{t_{\rmd}} = - f(\theta) \left(\frac{g(\theta)}{2}\right)^{1/3} \widetilde{t},
    \label{eq:timeScale}
\end{equation}
where $t_{\rmd} = \rho_0 R_0^2 / D c_s (1-H_r)$ is a diffusive evaporation time scale and $R_0 = \left(3V_0/4\pi\right)^{1/3}$ the volume-equivalent initial droplet radius. Eq. [\ref{eq:timeScale}] holds for early times $\widetilde{t}~<~0.7$ and collapses the data as can be seen in Fig. \ref{fig:allRH}.

\begin{figure*}
 	\centering
 	\includegraphics[width=0.9\columnwidth]{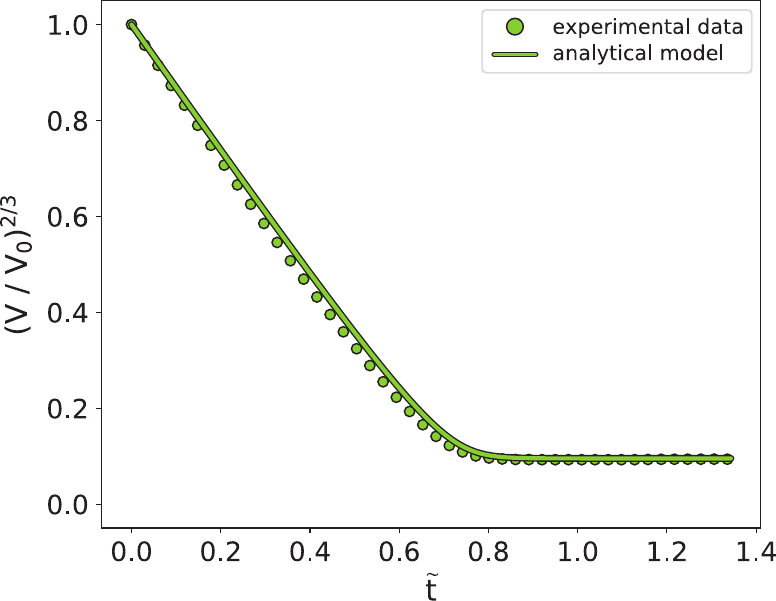}
 	\caption{\textcolor{black}{
 	Comparison between experimental data and analytical model. The graph shows the normalized droplet volume $(V/V_{0})^{2/3}$ with $V_{0}~=~V(t=0)$, against the normalized time $\widetilde{t}$, as also shown in Fig. \ref{fig:allRH}(b). As denoted in the legend, experimental data is shown with symbols, the analytical model with a line. The symbols for the experimental data were chosen in accordance with the other plots in this study. We present an example corresponding to a relative humidity 0.81. The model yields very good agreement with the experimental data for all the relative humidities explored, and in both limiting regimes: the diffusion-limited evaporation regime and the final stable volume.
 	}
 	}
 	\label{fig:ModelAgreement}
 \end{figure*}

\subsection*{Mass and salt content of stable liquid droplets at high relative humidities}

The mass of the final object, regardless of whether it consists of a dried object or rather a stable liquid droplet, is of crucial importance for two main reasons: first, the final mass will determine the sedimentation speed, i.e. it determines for how long the droplet will remain suspended in air. Secondly, the solute concentration (and most importantly, the concentration of salts) in a respiratory liquid droplet is crucial for the \textcolor{black}{inactivation of potentially present virions \cite{Yang2011AirborneDynamicsIAV}.} 

\begin{figure*}
 	\centering
 	\includegraphics[width=\columnwidth]{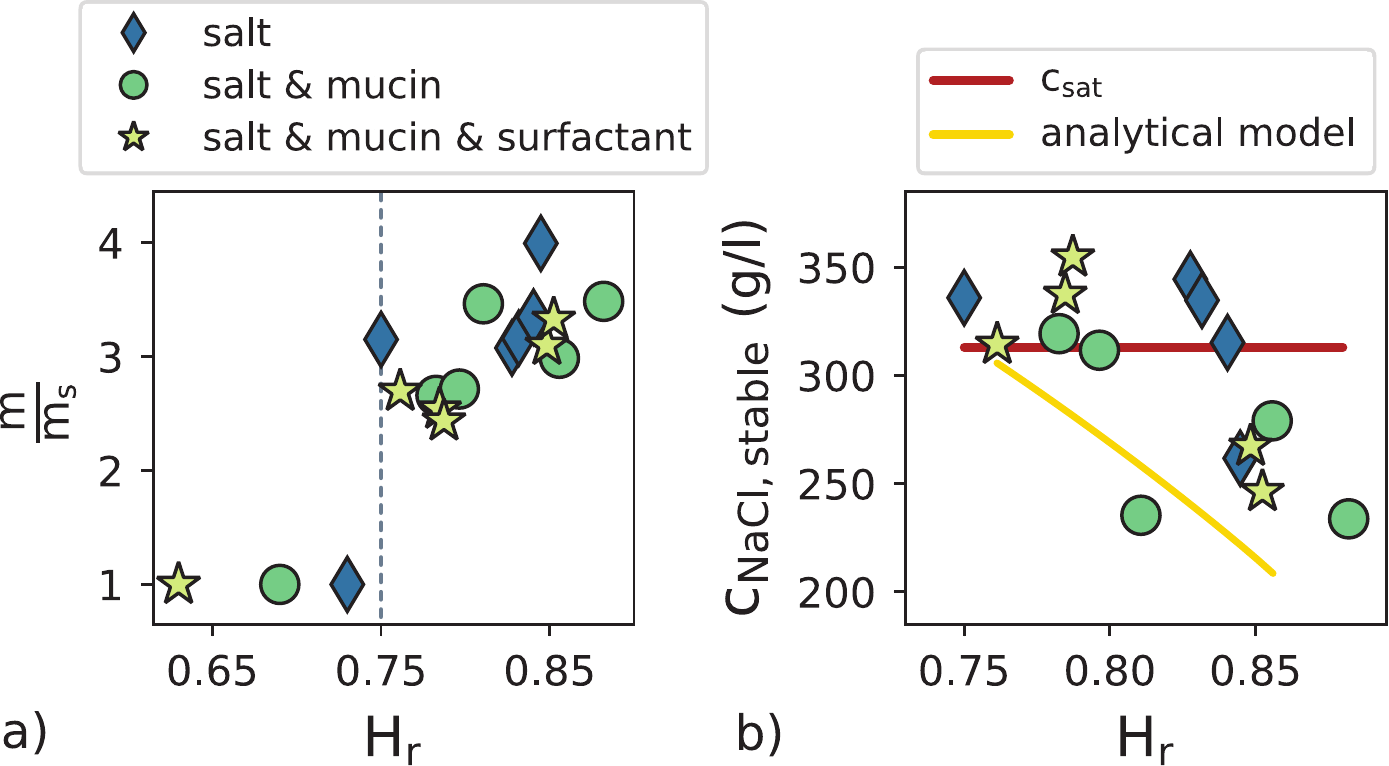}
 	\caption{\textcolor{black}{
 	(a) Mass of the remaining object with respect to the mass of non-volatile solute. Droplets at relative humidities below around 75\% dry out completely, leaving only a dry residue behind. Therefore, by definition, $m/m_s \equiv 1$ for those below the deliquescence limit (dashed vertical line).
 	Droplets at relative humidity above this value reach a stable liquid volume at a certain solute concentration, therefore $m/m_s>1$ to account for the liquid mass.
 	(b) Concentration of sodium chloride in solution in stable liquid droplets. The mass of sodium chloride is computed based on the initial concentration in \emph{solution}. The concentration $c_{NaCl, stable}$ denotes the mass of sodium chloride per unit volume in the stable drop.
 	The red line in this plot corresponds to the saturation concentration at $20^{\circ}$C: 313 g/L. Note that this concentration is given as mass per volume of solution in a stable drop. The golden line denotes the analytical model, discussed in the text. The droplets tend to stabilize at lower salt concentrations as the relative humidity increases.
 	}
 	}
 	\label{fig:SaltMassConc}
\end{figure*}

\textcolor{black}{
Figure \ref{fig:SaltMassConc}(a) shows the ratio between the mass of the stable droplet volume $m$ (or the dried residue, respectively) and the mass of the added, non-volatile solutes $m_s$. For droplets below the critical relative humidity of 75\%, which end up in a dry state, we assume that all solvent has evaporated and therefore $m/m_s$ takes the value 1 by definition. However, for those droplets evaporating at relative humidities above 75\%, which end up in a stable salty liquid droplet, the fraction of water left within the stable salty liquid droplet $m/m_s$ is not trivial to determine experimentally, but can be estimated analytically.} \textcolor{black}{We make use of our analytical model and of tabulated salt solution densities \cite{lide2005crc} to solve for the water mass fraction, given a known final volume and a known initial mass of salt \textcolor{black}{(and other added solutes)} within the droplet.} As the relative humidity increases, the liquid phase reaches equilibrium at lower and lower salt concentrations, since the equilibrium vapor pressure gets smaller and smaller. \textcolor{black}{Therefore, the salt concentration also stays below the saturation level % at the same temperature 
that is reached for $H_{r}~\leqslant ~0.75$ (313 g/L at $20^{\circ}$C), as it is shown in Fig. \ref{fig:SaltMassConc}(b)}.

\subsection*{Droplet dry-out and resulting structure}

For relative humidities below 75\%, the water completely evaporates and a dry object is left behind. Interestingly, all droplets collapse into three-dimensional structures with a shape that strongly depends on its components, and to a lesser extend on the humidity. A great advantage of evaporating the droplets in such a controlled way is that we can analyze the intact remains under a scanning electron microscope. 

\begin{figure*}%[h]
\centering
\includegraphics[width=1\textwidth]{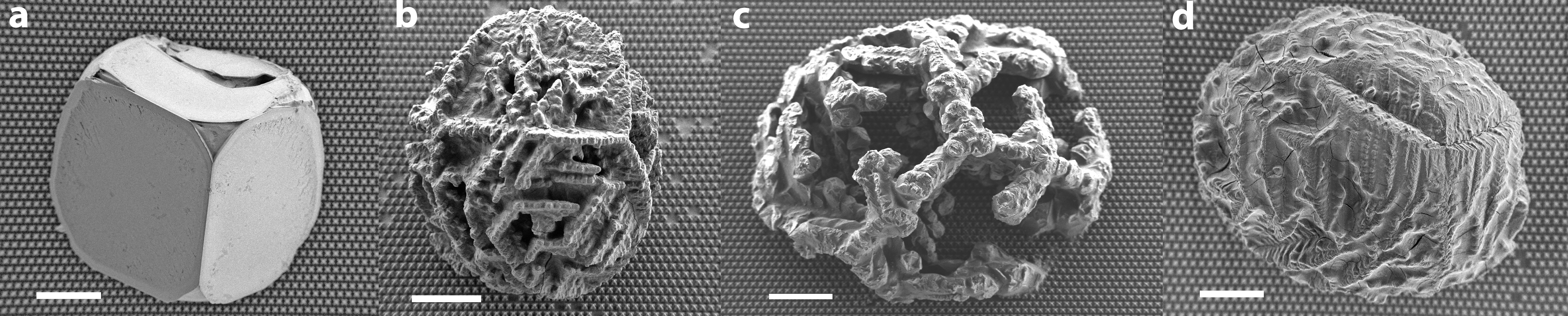}
\caption{Scanning electron micrographs taken from the remains of dried droplets containing (a) sodium chloride (initial droplet volume 3.5 $\upmu$L), (b) sodium chloride and mucin (initial droplet volume 1.4 $\upmu$L) and (c and d) sodium chloride, mucin and surfactant (respiratory-like droplets with initial volume 1.6 and 2.2 $\upmu$L respectively). All scale bars correspond to 100 $\upmu$m. The micrographs are taken facing the superhydrophobic substrate \cite{Berenschot2013fractalfabrication,seyfert2021evaporation} where the droplets are deposited. % further discussion comparing their sizes will depend on the initial droplet sizes. The salt droplet was 3.5 µL... 
More micrographs from other samples can be found in the supplementary materials.}
\label{fig:sem}
\end{figure*}

Figure \ref{fig:sem} shows the remains of droplets with different contents. In all cases, the droplets are evaporated until they reach stability (for $H_r \ge 0.75$) and then they are slowly dried by reducing progressively the ambient humidity. Fig. \ref{fig:sem}(a) shows the remains of a droplet containing only sodium chloride. The result is a single crystal-shaped remain\textcolor{black}{, with a final density which is close to that of crystallized sodium chloride. Therefore the salt residue appears to be massive and not porous.}. 

In Fig. \ref{fig:sem}(b) we observe the remains of a droplet containing salt and mucin. We can see that the presence of the protein disturbs the sodium chloride crystallization and forms several small-scale crystals, nucleated all over the interface.  
When all three solutes are added (sodium chloride, mucin protein and surfactant DPPC, in Fig. \ref{fig:sem}(b) and (c), the remains take different shapes and forms, without any apparent dependence on the process. They  either turn into an open structure as in Fig. \ref{fig:sem}(c), or rather closed by the protein/surfactant as in Fig. \ref{fig:sem}(d). Whenever salt and mucin are present, we always observe a "bone-like" structure made of sodium chloride crystals, coated by mucin protein and/or DPPC surfactant. All remains are three-dimensional, and keep an almost spherical shape, which is most likely due to the accumulation of non-volatile material at the liquid-air interface of the droplet.

This is a surprising phenomenon since a quick calculation can show that the time scale for the Brownian motion of, for example, a sodium ion is much shorter than the evaporation time scale. However, it can be shown that a salt concentration gradient can form in an evaporating droplet due to the non-linear dynamics of the receding interface \cite{gregson2018drying}, and a similar phenomenon has been shown to occur with colloids, yielding core-shell structures \cite{seyfert2021evaporation}. Consequently, it is not surprising that mucin and surfactant accumulate at the interface forming a sort of film, which could wrap (and protect) any other non-volatile content.
The compactness of dried respiratory-like droplets is quite low, yielding rather holey structures.
Interestingly, such dried shell microstructures would quickly rehydrate if the enviromental humidity is increased above the deliquescence limit ($H_{r}~>~0.75$) due to the presence of salts \cite{gregson2018drying,olsen2006single}.

Other studies in the literature using sodium chloride solution droplets on heated superhydrophobic substrates reported different structures \cite{shin2014igloos} and other exotic effects \cite{Noushine2020selflift} as self-lifting, which have not been observed in our case, probably due to the minimal contact with the solid substrate and the absence of deliberate heating.

\section*{Discussion}

% Discuss the implications of our findings in the context of the survival of enveloped viruses under such conditions (room temperature of 20 $^\circ$C and this range of humidities). \textcolor{red}{[Avoid discussing anything about transmission or infection probability.]}
In previous sections we have described how respiratory-like droplets behave under room temperature of $20^{\circ}$C and different relative humidities, in conditions that resemble the behavior of aerosol droplets suspended in air. We have identified three different regimes, each of which connected to a range of relative humidities and with different implications for the droplet's fate. In regime I, for a relative humidity $H_{r}~<~0.75$, the liquid phase evaporates completely and a dry object remains. 
In regime II, for relative humidities in the upper vicinity of approximately 75\%, the solution's vapor pressure reaches a critical value below atmospheric pressure and therefore the droplet volume remains stable at a maximum salt and solute concentration. As the humidity increases further, we reach regime III, in which the droplet volume remains stable at larger volumes, the salt and solute concentration is lower when compared with the previous regime, but the total mass is larger due to the larger amount of liquid water. 

We would like to discuss the implications of these three regimes on the survival of an enveloped virion particle, under the light of recent studies on the topic \cite{morris2021elife,lin2019humidity,Survival2020Marr} 
that have reported an U-shaped dependence of the activity of enveloped viruses with the ambient relative humidity.
%that have reported a U-shaped dependence of enveloped virions activity as a function of the relative humidity. 

In the following paragraphs we will work under the assumption that the presence of sodium chloride in solution challenges the survival of an enveloped virion particle, due to the osmotic pressure difference across its lipid membrane.
Enveloped biological systems are subjected to osmotic stress during drying processes and in high osmotic strength solutions. 
Most microorganisms, as well as human/animal/plant cells, maintain an osmotic pressure balance by the synthesis of certain \emph{osmoprotectant} molecules \cite{Osmoregulation1984}. However, enveloped viruses lack regulatory water channels or osmoprotectant molecules, which makes them more vulnerable to osmotic damage. Interestingly, virus-like liposomes are also used as vectors in common jabs as the flu vaccine, so there is also currently interest in their preservation under different environmental conditions \cite{choi2015fluvaccine}. Other factors like temperature and pH will surely have an effect on the viral activity, but our study focuses in the role of ambient humidity. We identified the three, humidity-dependent regimes and will now discuss the {osmotic environment} for each of the regimes.

% regime I 
% We can then discuss what would be the \emph{osmotic environment} on each regime. 
In regime I, sodium chloride crystallizes once a saturation concentration is reached, which reduces the amount of dissolved salts in solution. Despite of a salt crystal being a rather dry environment, enveloped viruses are known to be able to survive long periods of time in dry conditions. This was already known for the vaccinia virus \cite{collier1954preservation}, and actually spray drying is currently used to preserve vaccines based on viral liposome vectors \cite{vaccinedrying2021}. As we have shown in Fig. \ref{fig:sem}(b)-(d) and discussed in the previous section, the dry remains of a respiratory-like droplet adopt a shell structure that could actually serve as protection from the environment for any other element within the droplet, such as pathogens. 
In our model experiments, the initial droplet size is rather large, but a typical aerosol droplet size will be in the range below 20 $\mathrm{\mu m}$ \cite{Morawska2009,bourouiba2014violent,Bourouiba2021}, yielding dry remains in the range below 10 $\mathrm{\mu m}$. Consequently, such dry nuclei will easily remain in suspension in air until hitting a solid substrate or a host. As we have discussed in previous sections, if the dry nuclei is introduced into an environment with humidity above the deliquescence limit ($H_r~>~0.75$), it would quickly rehydrate and it would be able to disperse its content, for example into the respiratory track of a potential host.
Recent studies have shown how influenza A virus can be transmitted from a carrier to a receiver in the form of aerosolized fomites \cite{asadi2020influenza}. Although the study has only been tested on animal models with influenza A virus, it confirms that is a possible way of transmission for other enveloped viruses, this discovery confirms the extraordinary survival capabilities that enveloped viruses have in dry conditions.

% regime II
In regime II, i.e. at a relative humidity of around 75\%, a respiratory droplet containing an active enveloped virus would remain stable at a salt concentration close to saturation, which might potentially be unbearable for the viral envelope, according to recent results on enveloped virus vaccines \cite{choi2015fluvaccine}. 
{Animal and plant cell membranes (as well as some bacteria) are able to regulate their osmotic pressure difference by the exchange of ions through the Na$^+$-K$^-$ pump, which ultimately regulates the cell's volume \cite{BAUMGARTEN2012261}. Viral membranes do not have such a mechanism, and therefore are sensitive to dramatic volume changes depending on the osmotic pressure changes. Let us assume that the viral cytoplasm (not the capsid, which can sustain higher pressures \cite{Cordova2003osmotic}) is at rest under isotonic conditions in the human saliva. The amount of moles of solute that can potentially contribute to an osmotic pressure is typically known as Osmoles (Osm). Human saliva reaches osmolarity values of 279$\times10^{-3}$ Osm/L. In regime II, the osmolarity in the droplet increases to a remarkable value of approximately 12 Osm/L (highly hypertonic conditions), occurring at the deliquescence limit of about 75\% relative humidity, which we can translate into pressures by applying \emph{van't Hoff's} law \cite{BAUMGARTEN2012261}, yielding pressures of 1.47$\times10^7$ Pa, approximately 147 atm. Enveloped viruses as the influenza shrink irreversibly and become inactive within seconds under hyperosmotic stress differences of only 2.4 Osm/L (by NaCl) \cite{choi2015fluvaccine}; this is approximately 1/5 of the hyperosmotic stress that would be achieved in a salt-saturated droplet.}
A droplet at the deliquescence limit would therefore be the deadliest regime for an enveloped virion of all those covered in this study. Since the droplet would be rather small, it will manifest in an almost homogeneous and saturated salt concentration over its whole volume \cite{gregson2018drying}.
{Unfortunately, with the data at hand it is difficult for us to determine precisely the relative humidity range in which viruses would suffer from this mechanism we propose. From Fig. \ref{fig:SaltMassConc}(b) we show that the salt concentration in the stable droplets decreases rapidly with increasing relative humidity beyond 75\%, and the hyperosmotic stress across an unregulated lipid membrane for a pathogen inside droplet at 90\% would be approximately 5 Osm/L (about 60 atm of pressure), which would in principle be a hostile environment for an enveloped influenza virus according to previous studies \cite{choi2015fluvaccine}. We are however not in a position to speculate further. Viral activity assays \cite{morris2021elife,Survival2020Marr} should be performed under controlled conditions to be able to determine the viral activity decay with more precision.
Note that the lower limit of regime II is also not well determined, since a droplet slightly below the deliquescence limit (e.g. $H_r$ = 0.7), in the process of being volatilized, will also experience for a certain amount of time high sodium chloride concentrations, which could be also lethal for an enveloped virus. \textcolor{black}{Moreover, we must take into account that an airborne droplet can remain in a metastable liquid state with a salt concentration larger than the saturation one for a long time at relative humidities larger than the deliquescence one \cite{gregson2018drying}. This means that the harshest conditions for the virus may actually be observed, not at this relative humidity, but at an intermediate humidity between the deliquescence and the efflorescence points.} If the upper limit of regime II is defined by the amount of osmotic pressure necessary to damage an unregulated lipid membrane irreversibly, the lower limit is defined by the amount of time that is required to disrupt the virion at sufficient osmotic pressure, before salt crystallization kicks in, reducing the amount of salt in solution.}

% regime III
As the relative humidity further increases, we eventually enter regime III, in which the droplet stabilizes at larger volume with lower salt concentration. Additionally, in a larger droplet, the diffusion of the solutes might not be enough to maintain a homogeneous salt concentration over the whole droplet, which typically results in higher concentration in the vicinity of the interface, and a lower concentration in the droplet's bulk \cite{gregson2018drying}. This concentration gradient within the droplet can result in regions with rather low salinity in solution, and therefore with higher survival probability for an enveloped virus. 

These results give physical support to several studies that suggest that enveloped viruses have a higher tendency to survive at extreme values of humidities \cite{morris2021elife,Survival2020Marr,vejerano2018physico,lin2019humidity}, while their survival decreases to a minimum at intermediate values. Interestingly, the value of the humidity in which the minimum viral activity is found lies in the range between 50\% \cite{Survival2020Marr} and 65\% \cite{morris2021elife}, while according to our results, {this range of humidities should lay somewhere above those reported values, rather around 65\% and 85\%, with an expected minimum of viral activity in the vicinity of the deliquescence limit (relative humidity around 75\%),} \textcolor{black}{or somewhat smaller, if we consider the possibility that drops survive in the metastable supersaturated liquid state mentioned above.}

The range of relative humidities in which a minimum in the viral activity is found is crucial. Unfortunately such studies are complex and tedious, and the number of relative humidities explored is typically very limited to have precision \cite{morris2021elife,Survival2020Marr}. Droplets in such studies are evaporated inside large rotating drums at controlled humidity, measured with standard humidity sensors in one single location. We argue here that the humidity reported in those experiments cannot be directly compared with the far-field humidity experienced by an individual droplet, as we report in our work.

In our experiments, the reported relative humidity is measured by fitting the non-linear evaporation rate to our evaporation model of a single droplet, which has given us excellent results. In every experiment, the temperature and humidity of the chamber is registered using a standard humidity and temperature sensor SHT75 (see the schematic of the set-up in Fig. \ref{fig:1}(a) for its location) during the whole evaporation process to ensure that no abrupt changes occur.

On the other hand, experiments with aerosols are typically done in closed and rotating systems, and the humidity gradient experienced by individual droplets is heavily influenced by neighboring droplets depending on the location within the bounded system, or within the aerosol itself, as recent direct numerical simulations have predicted \cite{chong2021extended}. Connected to this, although temperature effects have not been discussed in our study, evaporative cooling might lower the saturation concentration and consequently reduce the effective deliquescence critical humidity.

For an isolated drop in an large air volume, the latent heat absorbed to evaporate the liquid is not able to reduce substantially the temperature of the surroundings. However, in a dense aerosol, the amount of air available per droplet may be small enough for evaporative cooling to lower its temperature, which would translate into the droplet evaporating in an environment effectively colder than the far-field air temperature outside the aerosol. However, such effect in isolated water droplets, as those considered here, is not significant to explain the difference between the value of the relative humidity for which we find maximum salinity in a droplet and the value of the relative humidity for which a minimum viral activity has been found in other studies \cite{Survival2020Marr,morris2021elife}. Nevertheless, the value found in our experiments for regime II, in which the salinity of the solution would be the highest and therefore the environment would be lethal for an enveloped virus, is in agreement with the well-known deliquescence point for salt solutions \cite{gregson2018drying}, which also agrees with our model based on basic principles.

Interestingly, a similar U-shape curve has been found when studying the seasonality of influenza infections across different climates around the world. The conclusions of studies like that of Tamerius et al. \cite{tamerius2013influenza} is that there are two extreme environmental conditions associated with seasonal influenza epidemics: a dry (and cold) and a humid (and rainy). By computing correlations for the different average humidities of each season, Tamerius et al. \cite{tamerius2013influenza} reproduce an influenza infection probability vs humidity curve with a remarkable U-shape, having the minimum values in approximately 75\% relative humidity; as the values found here for the deliquescence limit. However, we must be careful about this argument, as most of the infections take place indoors, where the temperature and humidity conditions in general differ from the ambient ones \cite{Bov2021Humidity}.

Before concluding this section we must point out that there may exist other mechanisms, besides osmotic pressure, that contribute to the inactivation of a virus in high salinity environments. For instance, when in aqueous solution, chlorine is also an effective disinfectant agent \cite{HaasEnvSciTechnol1980}.  An alternative effect that may inactivate the virus in the drop is the presence of free chlorine. Chlorine may chemically degrade both the virus binding sites as well as its genetic material. The presence of free chlorine in the sodium chloride solution is another mechanism by which the virus is inactivated in an evaporating droplet.

\textcolor{black}{Our analytical model could be easily implemented in state-of-the-art numerical simulations of aerosols \cite{chong2021extended}. Which would dramatically extend the predicted lifetime of respiratory droplets in humid environments. 
Combined with estimations of the viral decay in respiratory droplets (due to the presence of salt), such numerical models would enable, in the future, the accurate estimation of the percentage of active virions contained in aerosols as well as the prediction of their spreading in different situations.}

\subsection*{Methods\label{subsec:methods}}

The droplets in this study evaporate inside a closed (non-pressurized) chamber, which minimizes environmental influences on the droplets. See Fig.\ref{fig:1}(a) for a schematic of the set-up. We use a humidity controller (HGC 30 from DataPhysics) to set the humidity in the chamber. To reach very low relative humidity values, we make use of the hygroscopic behaviour of magnesium chloride and place ``water traps'' inside the chamber. This water traps are little vessels filled with dry magnesium chloride. This method leads to humidity values below 0.15. The values for humidity and temperature within the chamber are monitored throughout all experiments. The superhydrophobic substrate is placed in between a CCD camera (Ximea MQ013MG-ON) and a light source, to capture a shadowgraphic side view of the experiment. 

We deposit a droplet on top of the superhydrophobic substrate with a threaded-plunger syringe and a tapered fused silica capillary (glass syringe from Hamilton, Model 1750 LT Threaded Plunger SYR, silica capillary from Polymicro Technologies). The nominal external diameter of the capillary is 360 $\upmu$m. 
To avoid any sudden impact of the droplet on the substrate, we slowly increase its volume with the threaded plunger, while the droplet already touches the substrate. By retracting the capillary upwards, the droplet, now having reached volumes between 2 and 3 {$\upmu$l}, detaches from the capillary and gently comes to rest on top of the substrate.

The evaporation process is captured with frame rates between 0.1 to 1 frames per second, depending on the expected duration of the experiment, and therefore on the humidity set during the experiment. From the droplet images \textendash{} see an example shown in Fig. \ref{fig:1}(b) \textendash{} we extract contact angles and droplet volumes as functions of time.

\subsection*{Materials\label{subsec:materials}}

For this study, we use four different kinds of solutions, distinguished by the number of components: (a) pure water, consisting of a single component (1C), (b) salt solution (2C), which comprises the two components of water and sodium chloride, (c) three-component solution (3C), which additionally contains mucin, and (d) the respiratory fluid-like solution (4C), where surfactant is added to the 3C solution.
We recreate those solutions, based on what has been used in other works before \cite{morris2021elife,vejerano2018physico}.

As foundation for all solutions, we use purified water (Milli-Q\textregistered ~IQ 7000 Water Purification System). The organic components are purchased from Sigma-Aldrich. In all solutions, they have the constant initial concentration of 9 g/L (sodium chloride), 3 g/L (mucin from porcine stomach) and 0.5 g/L (1,2-Didodecanoyl-sn-glycero-3-phosphocholine, in short ``{DPPC}'').
% 1,2-Didodecanoyl-sn-glycero-3-phosphocholine 
% Description: surfactant in powder, biocompatible
% Amount: 100 mg 
% https://www.sigmaaldrich.com/catalog/product/sigma/p1263
% Mucin from porcine stomach 
% Description: protein solution in powder, biocompatible			
% Amount: 10 g
% https://www.sigmaaldrich.com/catalog/product/sigma/m1778
We mix the solutions on a magnetic stirrer, with minimal heating, in a covered glass beaker to avoid evaporation of the solvent during the mixing. Salt and DPPC dissolve fairly quickly in water under stirring, the mucin takes roughly four times as long to fully dissolve. Before we do experiments, we vigorously mix the solutions on a vortex mixer, and also mix them on a magnetic stirrer again, to dissolve potentially precipitated mucin. After mixing, we centrifuge the solution for 10 minutes and extract the supernatant for our experiments. This last step ensures that no precipiated, agglomerated solids contaminate our experiments.

\subsection*{Substrate\label{subsec:substrate}}

The droplets evaporate on superhydrophobic, microfabricated $\rm{SiO_{2}}$ substrates. The surface is decorated with fractal-like microstructures \cite{Berenschot2013fractalfabrication, FractalEngineering}. A classic vapour deposition protocol renders the substrate superhydrophobic, by depositing a layer of Fluoroctatrichlorosilane (FOTS) on top of the microfabricated structures. The substrates display a static contact angle of $\mathrm{\geqslant 155^{\circ}}$ for water droplets, and roll-off angles as low as $1.5^{\circ}$.

\subsection*{Acknowledgments}
The authors would like to thank Erwin J. W. Berenschot and Niels Tas for supplying the superhydrophobic substrates used in this paper. AM and CS acknowledge financial support from the European Research Council Starting Grant program, project number 678573. JRR acknowledges funding from the Spanish Ministry of Economy and Competitiveness/Agencia Estatal de Investigación through grant DPI2017-88201-C3-3-R, partly funded through European funds. DL acknowledges discussions with Mariette Knaap and funding from the Netherlands Organisation for Health Research and Development (ZonMW), project number 10430012010022, and from the European Research Council Advanced Grant program, project number 740479.

\end{multicols}

% Bibliography
% \bibliography{EvaporResp.bib}

\end{document}